\documentclass[aps,prl,twocolumn,showpacs,preprintnumbers,amsmath,amssymb,floatfix,nofootinbib,letterpaper]{revtex4}

\usepackage{graphics}
\usepackage{epsfig}
\epsfclipon

\usepackage{stmaryrd}

\def\Box{\boxempty}

\newcommand{\nco}{\newcommand}
\nco{\beq}{\begin{equation}} \nco{\eeq}{\end{equation}}
\nco{\beqa}{\begin{eqnarray}} \nco{\eeqa}{\end{eqnarray}}

\def\be{\begin{equation}}
\def\ee{\end{equation}}

\def\baray{\begin{eqnarray}}
\def\earay{\end{eqnarray}}

\nco{\sss}{\scriptscriptstyle} \nco{\dphi}{\varphi}
\nco{\lsim}{\mbox{\raisebox{-.6ex}{~$\stackrel{<}{\sim}$~}}}
\nco{\gsim}{\mbox{\raisebox{-.6ex}{~$\stackrel{>}{\sim}$~}}}

\def\IK{\relax{\rm I\kern-.20em K}}
\def\IM{\relax{\rm I\kern-.20em M}}

\def\lsim{\mbox{\raisebox{-.6ex}{~$\stackrel{<}{\sim}$~}}}
\def\gsim{\mbox{\raisebox{-.6ex}{~$\stackrel{>}{\sim}$~}}}
\def\sss{\scriptscriptstyle}

\def\done{\delta^{(1)}}
\def\dtwo{\delta^{(2)}}
\def\sH{\mathcal{H}}

\def\grad{\vec{\nabla}}

\def\L{\mathcal{L}}

\def\bea{\begin{eqnarray}}
\def\eea{\end{eqnarray}}

\newcommand{\2}{\frac{1}{2}}

\newcommand{\LF}{\left(}
\newcommand{\RF}{\right)}

\def\done{\delta_1}
\def\dtwo{\delta_2}
\def\sH{\mathcal{H}}
\def\grad{\vec{\nabla}}
\def\lap{\vec{\nabla}^2}
\def\Vint{V_{\mathrm{int}}}
\def\Uint{U_{\mathrm{int}}}

\def\L{\mathcal{L}}
\def\Jint{J_{\mathrm{int}}}
\def\Jinttwo{J_{\mathrm{int,2}}}
\def\Fbox{F\left(\frac{\Box}{m_s^2}\right)}

\def\Gbox{\Gamma\left(\frac{\Box}{m_s^2}\right)}

\def\Gomega{\Gamma\left(-\frac{\omega^2}{m_s^2}\right)}

\begin{document}

\title{Nonlocal Inflation}

\author{Neil Barnaby}

\affiliation{Canadian Institute for Theoretical Astrophysics,
University of Toronto, 60 St.\ George St.\, Toronto, Ontario M5S 3H8 Canada\\
Email: \emph{barnaby@cita.utoronto.ca} }



\vspace{-3mm}

\begin{abstract}
We consider the possibility of realizing inflation in nonlocal field theories containing infinitely many derivatives.  
Such constructions arise naturally in string field theory and also in a number of toy models, such as the $p$-adic string.
After reviewing the complications (ghosts and instabilities) that arise when working with high derivative theories we discuss
the initial value problem and perturbative stability of theories with infinitely many derivatives.  
Next, we examine the inflationary dynamics and phenomenology of such theories.  Nonlocal inflation can proceed
even when the potential is naively too steep and generically predicts large nongaussianity in the Cosmic Microwave Background.
\end{abstract}


\vspace{-3mm}

\maketitle

\section{Introduction}
\vspace{-6mm}

One of the most striking features of the dynamical equations of string field 
theory \cite{sft} (SFT) is the presence of an infinite number of derivatives (both time 
and space).  This nonlocal structure is also shared by many toy models such as the 
$p$-adic string \cite{padic_ref} and strings quantized on a random lattice 
\cite{random}.  Stringy motivations aside, higher derivative field theories
are interesting due to improved ultra-violet behaviour (infinite order 
theories can be finite \cite{pais}), novel dynamics and a rich 
mathematical structure.  Here we will be interested in the possibility that 
nonlocal theories derived from string theory could have played a role in 
the early universe \cite{pi}-\cite{ng2}.  Specifically, we will consider a class of nonlocal 
inflation models showing that slow-roll inflation proceeds even when the 
potential is naively too steep \cite{pi,lidsey}.  This observation suggests a novel means of 
circumventing the difficulty of finding suitably flat scalar field
potentials that has stymied attempts to obtain inflation from string theory.  Moreover, we 
will see that nonlocal inflation is a predictive scenario since these models 
generically lead to large nongaussianity in the Cosmic Microwave Background 
(CMB) \cite{ng1,ng2}.

Several non-inflationary cosmological applications of nonlocal theories
 - for example involving quintessence, brane-worlds and solitons - have been studied recently;
see \cite{new1}-\cite{new6} and references therein.

Despite these motivations, applications of nonlocal theories to 
physics are fraught with complications \cite{woodard}.  Such theories are often 
plagued by classical (Ostrogradski) instabilities \cite{ostro}.  In the 
case with infinitely many derivatives one may wonder whether there is any 
sense at all in which the theory
possesses an initial value formulation \cite{niky}.  We will see that 
the structure of the initial value problem (IVP) is
intimately related to the stability of the theory. 

\vspace{-4mm}
\section{Example Infinite Order Theories} 
\vspace{-4mm}

Before proceeding, we first provide some explicit examples of  nonlocal Lagrangians that can be derived from fundamental physics.  
The traditional, perturbative, approach to string theory involves constructing a field theory living on the two-dimensional world-sheet that is swept out 
by the motion of a string through the $D$-dimensional target space-time.  When the world-sheet theory is 
quantized the resulting string excitations are associated with a spectrum of particles propagating in target space.  

SFT, on the other hand, is a re-formulation of string theory directly in the target space, as an off-shell theory 
of the infinite number of fields associated with the states of the string.  
The Lagrangian that one derives for the tachyon field\footnote{The tachyon field $T(x)$ encodes the instability of the D$25$-brane in bosonic string theory
which carries no conserved charge and hence is expected to decay.  The decay process is described by the condensation of the tachyon field $T$.} in a level truncation is \cite{sft}
\begin{equation}
  \label{SFT_tac1}
  \mathcal{L} = -\frac{1}{g_s^2}\left[\frac{\alpha'}{2}(\partial T)^2 - \frac{T^2}{2} + \frac{K^{-3}}{3}\left(K^{-\alpha'\Box} T\right)^2\right]
\end{equation}
where $\alpha' = 1/m_s^2$ (with $m_s$ the string mass), the constant is $K = 4 / (3\sqrt{3})$, $g_s$ is the open string coupling and $\Box = -\partial_t^2 + \grad^2$
in flat space.

A second example is $p$-adic string theory \cite{padic_ref}.  In this case the world-sheet coordinates of the string are taken to be valued in the field of $p$-adic numbers, rather
than in the field of real numbers.  Under this replacement it is possible to exactly compute all amplitudes for the tachyonic state and then write down a field theoretic Lagragian that
reproduces these amplitudes.  The result is \cite{padic_ref}
\begin{equation}
\label{paction}
  \mathcal{L} = {1\over g_p^2}\LF-\2 T\, p^{-{\Box\over 2m_s^2}} T +{1\over p+1}T^{p+1}\RF
\end{equation}
where $p$ is a prime number and $g_p^2 = g_s^2(p-1)/p^2$.

\vspace{-4mm}
\section{Finitely Many High Derivatives}
\vspace{-4mm}

Before considering infinite order theories, let us briefly illustrate how 
higher derivatives lead to vacuum instability. Consider the theory
\begin{equation}
\label{Slw}
  \mathcal{L}_{\mathrm{LW}} = \frac{1}{2}\phi\left[\Box - \frac{\Box^2}{M^2} 
                                    - m^2 \right]\phi - \frac{g}{3!}\phi^3
\end{equation}
called the Lee-Wick model \cite{LW}.
The classical equation of motion for $\phi$ is fourth order in time 
derivatives and hence the solution $\phi(t,{\bf x})$ is uniquely specified by 
four initial data $\phi(0,{\bf x}), \dot{\phi}(0,{\bf x}), \cdots$
The presence of extra initial data suggests already that (\ref{Slw}) describes
more degrees of freedom that the usual Klein-Gordon theory.  This intuition 
is further supported by noting that the momentum-space propagator
$G(p) \sim i / (-p^2-p^4/M^2-m^2)$
has two poles (for $M^2 \gg m^2$ these are at $p^2 \cong -m^2$ and $
p^2\cong -M^2$).  We can make these degrees of freedom explicit by 
introducing an auxiliary field $\tilde{\phi}$ which is a Lagrange multiplier
\begin{equation}
  \mathcal{L}_{\mathrm{LW}} = -\frac{1}{2}(\partial\phi)^2 
  - \frac{m^2}{2}\phi^2 +  \tilde{\phi}\Box\phi 
  + \frac{M^2}{2}\tilde{\phi}^2 - \frac{g}{3!}\phi^3
\end{equation}
Removing $\tilde{\phi}$ from the Lagrangian using its equation of motion 
yields (\ref{Slw}).  Now we introduce a new field 
$\hat{\phi} = \phi + \tilde{\phi}$ and perform the rotation
$\hat{\phi} = \cosh\theta \chi_1 + \sinh\theta \chi_2$, $\tilde{\phi} = \sinh\theta \chi_1 + \cosh\theta \chi_2$ which, for appropriate choice of $\theta$, 
diagonalizes the mass matrix, giving:
\begin{eqnarray}
   \mathcal{L}_{\mathrm{LW}} &=& -\frac{1}{2}(\partial\chi_1)^2  - \frac{m'^2}{2}\chi_1^2+ \frac{1}{2}(\partial\chi_2)^2 + \frac{M'^2}{2}\chi_2^2 \nonumber \\
                             && \,\,\,\,\,\, - \frac{g'}{3!}(\chi_1 - \chi_2)^3 \label{Slw_final}
\end{eqnarray}
where $m'$, $M'$, $g'$ can be computed from $m$, $M$, $g$ as in \cite{LW-SM}.  
With the Lagrangian in the form (\ref{Slw_final}) it is clear that the theory 
(\ref{Slw}) describes two physical degrees of freedom: an ordinary scalar 
$\chi_1$ with mass $m'$ and a ghost-like field with wrong-sign kinetic term 
$\chi_2$ and mass $M'$.  The field $\chi_2$ will contributes to the Hamiltonian 
with negative kinetic energy, rendering the total energy density unbounded from 
below.  It follows that the theory (\ref{Slw}) is unstable, time evolution 
generically drives the system to become arbitrarily excited.

Notice that this pathology is, at its core, a purely classical one; the theory
 (\ref{Slw}) is already sick before quantization. The pathology associated 
with high derivative theories is
sometimes described as being a breakdown of unitarity in the quantum theory.  
This conclusion results from an incorrect quantization using a non-normalizable vacuum
state \cite{woodard2}.

Although we have illustrated the problem of instability in the context of 
(\ref{Slw}) the problem is actually very general and can be seen already at 
the level of the classical dynamics of a 
point particle with position $q(t)$.  Ostrogradski \cite{ostro} has 
shown that if the Lagrangian $L = L\left[q,\dot{q},\cdots,q^{(n)}\right]$ 
depends non-degenerately on $q^{(n)}(t)$ then the Hamiltonian is 
\emph{always} unbounded from below when $n > 1$ (strictly this theorem only
holds for $n$ finite).  Notice the connection between stability and the counting 
of initial conditions: if the Euler-Lagrange equation for $q(t)$ admits more than 
two initial conditions then the theory will be unstable.

\vspace{-4mm}
\section{Infinitely Many Derivatives}
\vspace{-4mm}

We have seen how the stability of a higher derivative theory is intimately 
tied to the counting of initial data.  In order to assess the stability of 
theories containing infinitely many derivatives we should first seek to 
understand the IVP for infinite order partial differential equations.  A 
mathematical analysis was undertaken in \cite{niky} using the formal
generatrix calculus of pseudo-differential operators (see \cite{niky2} for the generalization
to variable coefficient equations).  Here we simply 
state the relevant result.  Consider the nonlocal 
theory
\begin{equation}
\label{proto2}
   \mathcal{L} = \frac{1}{2}\phi H(\Box) \phi - V_{\mathrm{int}}(\phi) 
\end{equation}
where $V_{\mathrm{int}}$ contains only terms $\mathcal{O}(\phi^3)$ and higher.
Ignoring the interactions the field equation is $H(\Box)\phi = 0$.  In 
\cite{niky} it was shown that general solutions $\phi(t,{\bf x})$ admit 
$2N$ initial data where $N$ counts the zeroes of $H(z)$ by multiplicity.  
This result is easy to understand on physical grounds since the propagator 
$G(-p^2)\sim 1/H(-p^2)$ has $N$ poles.  Physically, each pole of the 
propagator should correspond to an excitation in the theory and hence one 
expects two initial data per physical state (these are the two free 
coefficients that are promoted to annihilation/creation operators on 
quantization).  This leads to the (perhaps) surprising result that infinite 
order equations need not admit infinitely many initial data.

To establish (in)stability we should 
actually compute the Hamiltonian.  We perform a multi-particle decomposition 
similar to what was done for the Lee-Wick model in (\ref{Slw_final}), following \cite{pais}.  
Assuming meromorphic $H(z)$ 
with $M$ zeroes $H(m_i^2)=0$ (the $i$-th zero being of order $r_i$) the 
Weierstrass factorization theorem ensures that we can write
\begin{equation}
\label{Hz}
  H(z) = \Gamma(z)\prod_{i=1}^M \left(z - m_i^2\right)^{r_i}
\end{equation}
with $\Gamma(z)$ having no zeroes (although it may have isolated poles).  
Note that the number of zeroes counted by multiplicity is $N \equiv \sum_i r_i$
with $i=1,\cdots, M$.  Introducing $M$ independent\footnote{By ``independent''
here we mean that each $\phi_i$ has its own $2r_i$ initial data.  The proof of
this statement is implied by the analysis of \cite{niky}.} fields
\begin{equation}
\label{phii}
  \phi_i \equiv \prod_{j\not= i} (\Box - m_j^2)^{r_j}\phi
\end{equation}
we can decompose the Lagrangian (\ref{proto2}) into a sum of interacting theories as
\begin{equation}
  \mathcal{L} = \frac{1}{2}\sum_{i=1}^{N}\eta_i \phi_i \Gamma(\Box) (\Box - m_i^2)^{r_i} \phi_i - V_{\mathrm{int}}  \label{decomp} 
\end{equation}
Setting aside, for a moment, the question of how the constants $\eta_i$ are computed, let us try to determine when (\ref{decomp})
can yield a bounded Hamiltonian, at least perturbatively.  Clearly a necessary condition is that each $\mathcal{L}_i = \frac{1}{2}\eta_i \Gamma(\Box) \phi_i (\Box - m_i^2)^{r_i}\phi_i$ 
should give a Hamiltonian $\mathcal{H}_i$ that is positive definite (otherwise inclusion of interactions will certainly destabilize the system).  In order to have
$\mathcal{H}_i \geq 0$ we certainly need $m_i^2$ real and $r_i=1$ for all $i$ \cite{pais}.  In this case $N = M$ and the coefficients $\eta_i$ are defined by
\begin{equation}
\label{eta1}
  \sum_{i=1}^N\frac{\eta_i}{(z-m_i^2)} = \frac{\Gamma(z)}{H(z)} \equiv \frac{1}{\bar{H}(z)}
\end{equation}
Each $\eta_i$ can be computed from the residue of the propagator at the pole $m_i^2$ as
\begin{equation}
\label{eta2}
  \eta_i = \frac{1}{2\pi i}\oint_{C_i} dz\, \frac{1}{\bar{H}(z)} = \frac{1}{\bar{H}'(m_i^2)}
\end{equation}
where the contour $C_i$ only encloses the pole at $z=m_i^2$.

Let us return to the problem of determining when $\mathcal{H}_i > 0$.  At the linearized level we can remove the factor $\Gamma(\Box)$ with a nonlocal field redefinition \cite{niky}.  
Now $\mathcal{H}_i$ is computed from $\mathcal{L}_i = \frac{1}{2}\eta_i \Gamma(m_i^2) \chi_i(  \Box - m_i^2)\chi_i$
where the factor $\Gamma(m_i^2)$ appears to ensure that the field redefinition $\phi_i \rightarrow \chi_i$ yields real-valued $\chi_i$.  Clearly, then,
$\mathcal{H}_i \geq 0$ implies that $\epsilon_i > 0$ for all $i$, where $\epsilon_i \equiv \eta_i \Gamma(m_i^2)$.  If $N = 1$ this condition is straightforward.
Suppose $N > 1$.  By construction $\bar{H}(z)$ is analytic so $\eta_i$ must flip signs.  If we want $\epsilon_i > 0$ for all $i$ then $\Gamma(m_i^2)$ must
also flip signs.  Since $\Gamma(z)$ has no zeroes, this can only happen if it has simple poles between each zero of $H(z)$.  We thus conclude that (\ref{proto2})
can be perturbatively ghost-free in two cases:
\begin{enumerate}
  \item A single zero, $N=1$, in which case the theory describes only one degree of freedom.
  \item Multiple zeroes $N>1$ with $H(z)$ non-analytic.  Here the theory describes many degrees of freedom.
\end{enumerate}
We note that this approach does not establish nonperturbative stability.  Indeed, there are known examples of single pole theories that are non-perturbatively unstable \cite{woodard}.  
In the multi-pole case one can easily find examples that are completely stable non-perturatively by integrating out local fields.  See \cite{mulryne} for a discussion of nonlinear stability of nonlocal theories.

\vspace{-4mm}
\section{Why Nonlocal Inflation?}
\vspace{-4mm}

Many string theorists and cosmologists have turned their attention to building and
testing stringy models of inflation in recent years; see \cite{jim} for a review.  The goals have been to find
natural realizations of inflation within string theory, and novel features which would
help to distinguish the string-based models from their more conventional field theory 
counterparts.

In most examples to date, string theory has been used to derive an effective 4D field theory
operating at energies below the string scale and all the inflationary predictions are made
within the context of this low energy effective field theory.  This is a perfectly valid approach to
string cosmology, however, the use of low energy
effective field theory tends to obfuscate the stringy origins of the model.  As a result, it is often very difficult
to identify features of string theory inflation that cannot be reproduced in more conventional models.

Thus, there is motivation to consider models in which inflation takes place at higher energy scales where
stringy corrections to the low energy effective action are playing an important role.  This is usually daunting 
since the field theory description should be supplemented by an infinite number of higher dimensional
operators at energies above the string scale, whose detailed form is not, in general, known.    Here 
we propose to take a small step in the direction of overcoming this barrier,
by considering a simplified models of string theory inflation based on the UV complete models discussed previously.  
Since nonlocality is ubiquitous in SFT it is interesting to consider
a broad class of nonlocal inflationary models.

\vspace{-6mm}
\section{Nongaussianity in the CMB}
\vspace{-4mm}

Nongaussianity in the Cosmic Microwave Background (CMB) is usually
characterized by the dimensionless nonlinearity parameter, $f_{NL}$, which is
derived from the bispectrum of the gauge invariant curvature perturbation, $\zeta$. More explicitly, we have
\begin{equation}
  \langle \zeta_{k_1} \zeta_{k_2} \zeta_{k_3}\rangle = (2\pi)^3
  \frac{6}{5} f_{NL} \sum_{i<j}P_{k_i} P_{k_j}  \delta(k_1+k_2+k_3)
\end{equation}
with $P_{k}$ the power spectrum.  In a free theory $f_{NL}=0$ while for a theory with only a cubic interaction
$\mathcal{L}_{\mathrm{int}} = g \phi^3$ one has $f_{NL} \propto g$ (for this heuristic argument we are 
neglecting gravitational perturbations).  Hence, the nongaussianity is a measure of the strength of interactions 
in the theory.   In the simplest models of inflation $f_{NL}$ is suppressed by slow roll parameters and hence one has $|f_{NL}| \ll 1$  \cite{maldacena}.  
On the other hand, if $|f_{NL}| \gsim 5$ is observed it would rule out the
minimal inflationary scenario and favour one of a relatively small number of
nonminimal constructions.  

The current limit on nongaussianity in the CMB is $-9 < f_{NL} < 111 $ for the WMAP 5-year data\footnote{This is assuming the ``local'' form of nongaussianity.  For the equilateral model
one has the slightly weaker constraint $-151 < f_{NL} < 253$.} \cite{WMAP5} and future missions are
expected to be able to probe $f_{NL}$ down to order unity.

\vspace{-4mm}
\section{Nonlocal Hill-Top Inflation}
\vspace{-4mm}

For convenience we re-write the Lagrangian (\ref{proto2}) in the form
\begin{eqnarray}
  \mathcal{L} &=& \gamma^4\left[ \frac{1}{2}\phi \Fbox \phi - V(\phi) \right] \label{L} \\
  U(\phi) &=& U_0 - \frac{\mu^2}{2}\phi^2 + \frac{g}{3!}\phi^3 + \cdots \nonumber
\end{eqnarray}
where $\gamma$, $m_s$ have dimensions of energy and $\phi$, $F$, $\mu$, $g$ are dimensionless.  Without loss of generality we set $F(0)=0$.
We seek inflationary solutions rolling away from the unstable maximum $\phi=0$.  For perturbative stability we demand that $F(z)$ can be cast in the form:
\begin{equation}
\label{F}
  \gamma^2 \left[ \Fbox + \mu^2 \right] = \Gbox ( {\Box} + \omega^2 )
\end{equation}
where $\Gamma(z)$ is an analytic function of the complex variable $z$ having no zeros at finite $z$ \cite{niky}.  Notice that 
$ \mu^2 = -F(-\omega^2 / m_s^2)$ so that, depending on $F(z)$, there may be a large hierarchy between $\omega/m_s$ and $\mu$.

\vspace{-6mm}
\subsection{Background Evolution}
\vspace{-4mm}

In \cite{ng1} the evolution of the classical background $\phi(t)$, $H(t)$ was solved for using an expansion in powers of $u \equiv e^{\lambda t}$:
\begin{equation}
  \phi(t) =  \sum_{r=1}^{\infty} \phi_r e^{r \lambda t}, \hspace{3mm} H(t) = H_0 - \sum_{r=1}^{\infty} H_r e^{r \lambda t}
\end{equation}
The solution is parametrized so that at $t \rightarrow -\infty$ the field sits at the unstable maximum $\phi=0$ and drives a de Sitter expansion with Hubble scale $H_0$.
The solutions are $\phi(t) \cong e^{|\eta| H_0 t}$, $H(t) \cong H_0 - H_2 e^{2|\eta| H_0 t}$
where $3H_0^2 = \frac{V_0}{M_p^2} = \frac{\gamma^4 U_0}{M_p^2}$ and $H_2$ is computed in \cite{ng1}.
We employ the slow roll parameter $\eta = -\frac{\omega^2}{3H_0^2}$ which is different from
the naive $\eta$-parameter one would infer from (\ref{L}) in a derivative truncation $\Fbox \cong F'(0)m_s^{-2} \Box$.
In particular, slow roll can proceed even when the naive $\eta$-parameter is very large.  We have verified this remarkable behaviour using a nonperturbative 
analytical formalism in \cite{pi} and in \cite{diffusing} it was verified using nonlinear numerical simulations.

It is convenient to also define a second slow-roll parameter $\epsilon \equiv -\dot{H} / H^2$.
As is typical of hill-top models, we have large hierarchy between the slow roll parameters $\epsilon \ll |\eta|$.  We can therefore consistently work to leading order in $\eta$
but zeroth order in $\epsilon$.

\vspace{-5mm}
\subsection{Cosmological Perturbations}
\vspace{-4mm}

To study perturbations we introduce the ``canonical'' field $\varphi$
\begin{equation}
\label{varphi}
  \varphi  = \gamma \, \Gbox^{1/2} \phi
\end{equation}
In terms of which (\ref{L}) takes the form
\begin{equation}
\label{L2}
  \L = \frac{1}{2}\varphi \left(\Box + \omega^2 \right)\varphi - V_0 - \Vint\left[\Gbox^{-1/2}\varphi\right]
\end{equation}
where $V_0 = \gamma^4 U_0$ and $\Vint\left[x\right] = \gamma^4 \Uint\left[x / \gamma \right]$. We employ the Seery et al.\ \cite{seery} formalism
to compute the bispectrum and hence our starting point is the canonical field equation
\begin{equation}
\label{EOM2}
  (\Box + \omega^2)\varphi = \Gbox^{-1/2}{\Vint'}\left[\Gbox^{-1/2}\varphi \right]
\end{equation}

We perturb the field as
\begin{equation}
  \varphi(t,{\bf x}) = \varphi_0(t) + \delta \varphi(t,{\bf x}) = \varphi_0(t) + \done\varphi(t,{\bf x}) + \frac{1}{2}\dtwo\varphi(t,{\bf x})
\end{equation}
where $\done\varphi$ is defined so that it obeys exactly gaussian statistics.  We employ uniform curvature gauge and introduce conformal time $\tau$ defined by $a d\tau = dt$. 
We denote derivatives with respect to $\tau$ by $f' = \partial_\tau f$ and define a conformal time Hubble scale $\sH = a' / a$.
At linear order the (\ref{EOM2}) gives 
\begin{equation}
\label{phi1}
\done\varphi'' + 2\sH \done\varphi' - \lap\done\varphi - a^2\omega^2\done\varphi  - 2\epsilon \frac{a^2 V}{M_p^2}\done\varphi = 0
\end{equation}
Since the nonlocal structure of $\Vint$ in the Lagrangian (\ref{L2}) does not appear at the linear level, the equation for $\done\varphi$ 
is precisely the same result that one would have in a standard (local) model.  In the limit $\epsilon / |\eta| \rightarrow 0$ the perturbation
$\done\varphi$ is an eigenfunction of the de Sitter space d'Alembertian and the solutions are well-known.  Computing the curvature perturbation
$\zeta$ and the power spectrum in the usual manner we have spectral index $n_s -1 \cong 2\eta = -\frac{2\omega^2}{3H_0^2}$ \cite{ng2} so that $n_s < 1$,
in agreement with \cite{WMAP5}.

In order to see the effect of nongaussianity we must consider the second order perturbation $\dtwo \varphi$.  All of the contributions coming from the left-hand-side of (\ref{EOM2})
appear already in the standard (local) theory and have been computed in \cite{malik}.  The only new contribution coming
from the nonlocal structure of (\ref{L2}) arises due to the term $\Vint\left[\Gamma^{-1/2}(\Box / m_s^2)\varphi\right]$ that appears on the right-hand-side
of (\ref{EOM2}).  At leading order in slow roll parameters, the equation for $\dtwo\varphi$ is
\begin{equation}
\label{phi2}
  \dtwo\varphi'' + 2\sH\dtwo\varphi' - \lap\dtwo\varphi -a^2 \omega^2\dtwo\varphi = \Jinttwo + F_2(\done\varphi) + G_2(\done\varphi)
\end{equation}
where the source terms $F_2$, $G_2$ are the same as those appearing in \cite{seery}.  The quantities $F_2$, $G_2$ were explicitly computed in terms of field perturbations in 
\cite{malik} and their contributions to $f_{NL}$ were derived in \cite{seery}.  The only new term in (\ref{phi2}) is $\Jinttwo$ which denotes
the leading contribution in a perturbative expansion of
\begin{equation}
\label{Jint}
  \Jint = -2a^2 \Gbox^{-1/2} \Vint'\left[\Gbox^{-1/2}\varphi\right]
\end{equation}
Plugging the solutions of (\ref{phi1}) into (\ref{Jint}) we find \cite{ng2} 
\begin{equation}
\label{Jint_final}
  J_{\mathrm{int},2} \cong - a^2 \frac{g \gamma }{\Gomega \Gamma(0)^{1/2}} \, \left(\done\varphi\right)^2
\end{equation}
This takes exactly the form which would arise from a cubic interaction of 
the form $\mathcal{L}_{\mathrm{int}} = -{c_V} (\done\varphi)^3$ where the effective cubic coupling is
$c_V \equiv\frac{g \gamma}{3!}\Gomega^{-1}\Gamma(0)^{-1/2}$.
Depending on $F(z)$ this effective coupling can be much larger than the naive coupling $g$ (in the same way that the effective mass $\omega$ can be much
smaller than the naive mass inferred in a derivative truncation). If $c_V \gg g$ then this source term leads to large nongaussianity analogous to the way that a 
large cubic coupling would yield a large bispectrum in local field theory.  However, the novelty here is that this large (effective) interaction does not spoil the 
slow roll dynamics of the background.

In \cite{ng2} the bispectrum and nonlinearity parameter are explicitly computed in terms of model parameters.  The resulting nonlinearity parameter is very close to
the ``local'' ansatz $f_{NL} = \mathrm{const}$, allowing nonlocal inflation to be observationally distinguished from other models that give large nongaussianity, such
as small sound speed constructions \cite{chen}.

\vspace{-5mm}
\subsection{Example: $p$-adic Inflation}
\vspace{-4mm}

We now focus on the model (\ref{paction}).  In this case the COBE normalization implies a relationship between $p$, $g_s$ and $n_s$ which, for reasonable
values of the spectral index, constrains $g_s^2 / p \ll 1$.  There are two qualitatively different regions of the parameter space.  For $g_s \ll 1$ and $p = \mathcal{O}(1)$
the higher derivative structure of (\ref{paction}) is not playing an important role in the dynamics\footnote{In the limit $p\rightarrow 1$ (\ref{paction}) collapses to a local field
theory.} and one has $f_{NL} \sim n_s-1$, as in a local theory.  On the other hand, for $p \gg 1$, $g_s \sim 1$ the higher derivative terms in (\ref{paction}) cannot be neglected
and large nongaussianity is generated.  This latter regime is presumably the most natural from the string theory perspective since $g_s \ll 1$ usually requires fine tuning.

The nonlinearity parameter on equilateral triangles depends only on $p$, $n_s \sim 0.96$ and the number of e-foldings from horizon crossing to the end of inflation, $N_e \sim 60$:
\begin{equation}
\label{pfNL}
  f_{NL}^{{\triangle}} = \frac{5(N_e - 2)}{24\sqrt{2}}|n_s-1|^2 
e^{-\frac{N_e}{2}|n_s-1|}\, \frac{1}{\ln p}\frac{p-1}{\sqrt{p+1}}
\end{equation}
This is observably large in the interesting region of parameter space $g_s \sim 1$.  For reasonable parameter values this prediction
is consistent with the claimed detection in \cite{detect}.

\vspace{-6mm}
\section{Conclusions}
\vspace{-4mm}

We have considered the possibility of realizing inflation in nonlocal field theories with infinitely many derivatives.  These models provide
a playground for studying string cosmology to all orders in $\alpha'$ and display a number of novel features, including observably large
nongaussianity and slow roll even when the naive potential is steep.  The latter possibility is tantalizing because it suggests a novel way of
circumventing the difficulty of finding flat potentials in string theory by using the higher derivative structure that arises naturally in SFT.
Note that $|f_{NL}| \gg 1$ is generic in the sense that the nongaussianity is large when the nonlocal structure of the theory is playing an important
roll in the dynamics.

I am grateful to the organizers of Theory Canada 4 for their hospitality and for arranging for a stimulating conference.  Thanks
in particular to R.\ Brandenberger for inviting me to speak.

\end{document}